\documentclass[a4paper,11pt]{article}
\usepackage{pos}
\usepackage{booktabs}
\usepackage{float}

\title{Hessian-free force-gradient integrators and their application to lattice QCD simulations}
\ShortTitle{Hessian-free force-gradient integrators}

\author*[a]{Kevin Schäfers}
\author[b,c]{Jacob Finkenrath}
\author[a]{Michael Günther}
\author[b]{Francesco Knechtli}

\affiliation[a]{Institute of Mathematical Modelling, Analysis and Computational Mathematics (IMACM), Chair of Applied and Computational Mathematics, Bergische Universität Wuppertal,\\
Gaußstraße 20, Wuppertal, Germany}

\affiliation[b]{Department of Physics, Bergische Universität Wuppertal,\\
Gaußstraße 20, Wuppertal, Germany}

\affiliation[c]{Department of Theoretical Physics, European Organization for Nuclear Research, CERN,\\
1211, Genève, Switzerland}

\emailAdd{kschaefers@uni-wuppertal.de}
\emailAdd{j.finkenrath@cern.ch}
\emailAdd{guenther@uni-wuppertal.de}
\emailAdd{knechtli@uni-wuppertal.de}

\renewcommand{\logo}{\relax}

\abstract{
We present initial results on Hessian-free force-gradient integrators for lattice field theories.
Integrators of this framework promise to provide substantial performance enhancements, particularly for larger lattice volumes where higher-order integrators demonstrate greater efficiency.
Numerical results demonstrate the superior efficiency of the proposed integrators compared to commonly employed non-gradient schemes, particularly due to enhanced stability properties.
It is anticipated that the advantages of the Hessian-free framework will become even more pronounced in nested integration approaches and for smaller fermion masses, where the numerical stability properties of the integrators become increasingly important.\medskip

\texttt{Preprint CERN-TH-2025-024}
}

\FullConference{The 41st International Symposium on Lattice Field Theory (LATTICE2024)\\
 28 July - 3 August 2024\\
Liverpool, UK\\}

\begin{document}
\maketitle

\section{Introduction}
\subsection{The Hybrid Monte Carlo algorithm}
A commonly employed method for simulating quantum field theories on the lattice is the Hybrid Monte Carlo (HMC) algorithm~\cite{duane1987hybrid}. 
In the molecular dynamics (MD) step of the HMC algorithm, a separable Hamiltonian system $\mathcal{H}(p,q) = \mathcal{T}(p) + \mathcal{S}(q)$ must be solved employing a time-reversible and volume-preserving geometric numerical integration scheme to satisfy the fundamental detailed balance condition. 
Considering gauge field simulations in lattice QCD on a four-dimensional lattice of size $V = T \times L^3$ with lattice spacing $a$, the Hamiltonian is defined as
\begin{equation}\label{eq:Hamiltonian}
\mathcal{H}([P],[U]) = \mathcal{T}([P]) + \mathcal{S}([U]), 
\end{equation}
where $\mathcal{T}([P]) = \tfrac{1}{2} \sum_{x,\mu} \mathrm{tr}(P_{x,\mu}^2)$ represents the kinetic energy and $\mathcal{S}([U])$ denotes the action.
Here, the links $U_{x,\mu}$, connecting the sites $x$ and $x + a\hat{\mu}$, are elements of the special unitary group $\mathrm{SU}(3)$, whereas the scaled momenta $iP_{x,\mu}$ are elements of the corresponding Lie algebra $\mathfrak{su}(3)$ of traceless and anti-Hermitian matrices.
Any element $P$ can be expressed as $P = p^i T_i$ where $T_i$ denotes the generators of the Lie algebra.
The linear differential operators $\boldsymbol{e}_i$ act on the Lie group elements $U$ as $\boldsymbol{e}_i U = -T_i U$. They are gauge-covariant generalizations of the vector fields $\partial/\partial q^i$.
Consequently, the solution to the Hamiltonian system \eqref{eq:Hamiltonian} can be formally expressed through the $t$-flow  
\begin{equation}\label{eq:t-flow}
    \exp(t(\hat{\mathcal{A}} + \hat{\mathcal{B}})), \quad \text{with}\quad \hat{\mathcal{A}} = p^i \boldsymbol{e}_i \quad \text{and} \quad \hat{\mathcal{B}} = -\boldsymbol{e}_i(\mathcal{S}) \frac{\partial}{\partial p_i}.
\end{equation}

\subsection{Splitting methods}
Thanks to the separability of the Hamiltonian, \emph{splitting methods}~\cite{mclachlan2002,blanes2024} enable the construction of explicit geometric numerical integration schemes satisfying the detailed balance condition. 
Particularly, splitting methods compute a numerical approximation to \eqref{eq:t-flow} by composing evaluations of the flows $\exp(t \hat{\mathcal{A}})$ and $\exp(t \hat{\mathcal{B}})$ that can be computed exactly:
\begin{align}
    \exp(a_j h \hat{\mathcal{A}}) (P,U) &= (P, \exp(-a_j hP)U)  & &\text{(link update),} \label{eq:link-update} \\
    \exp(b_j h \hat{\mathcal{B}}) (P,U) &= (P - b_j h \boldsymbol{e}_i(\mathcal{S}) T^i , U) & &\text{(momentum update)}. \label{eq:momentum-update}
\end{align}
This leads to numerical integration schemes of the form 
\begin{equation}\label{eq:splitting_method}
    \Phi_h = \mathrm{e}^{b_s h \hat{\mathcal{B}}} \mathrm{e}^{a_s h \hat{\mathcal{A}}} \cdots \mathrm{e}^{b_1 h \hat{\mathcal{B}}} \mathrm{e}^{a_1 h \hat{\mathcal{A}}}.
\end{equation}
As the momentum and link updates both define symplectic maps, the overall splitting method is symplectic (and thus volume-preserving) as a composition of symplectic maps. 
Moreover, if the composition is self-adjoint, the method is also time-reversible.

\subsection{Force-gradient integrators}
Given that the kinetic energy $\mathcal{T}$ is only quadratic, force-gradient integrators \cite{omelyan2003symplectic,Chin2000force,Kennedy2009FGI,Clark2011FGI} represent a promising extension of splitting methods by incorporating a specific commutator, known as the \emph{force-gradient term},
\begin{equation}\label{eq:fg_term}
    \hat{\mathcal{C}} = [\hat{\mathcal{B}}, [\hat{\mathcal{A}},\hat{\mathcal{B}}]] = 2 \hat{\mathcal{B}} \hat{\mathcal{A}} \hat{\mathcal{B}} = 2 \boldsymbol{e}^j(\mathcal{S}) \boldsymbol{e}_j \boldsymbol{e}_i(\mathcal{S}) \frac{\partial}{\partial p_i},
\end{equation}
into the computational process. 
The force-gradient term is solely dependent on the links, enabling its inclusion within the momentum updates \eqref{eq:momentum-update}. This results in a \emph{force-gradient step}
\begin{equation}\label{eq:fg-step}
    \exp(b_j h \hat{\mathcal{B}} + c_j h^3 \hat{\mathcal{C}}) (P,U) = (P - b_j h \boldsymbol{e}_i(\mathcal{S}) T^i + 2c_j h^3 \boldsymbol{e}^j(\mathcal{S}) \boldsymbol{e}_j \boldsymbol{e}_i(\mathcal{S}) T^i,U).
\end{equation}
As the force-gradient step can be considered a common momentum update applied to a modified Hamiltonian system with perturbed action, it still defines a symplectic map.
A comprehensive classification of force-gradient integrators, as well as non-gradient schemes (splitting methods), with up to eleven stages has been presented in~\cite{omelyan2003symplectic}, highlighting the efficiency of the force-gradient approach.
However, the force-gradient approach entails certain drawbacks. 
Firstly, evaluating the force-gradient term \eqref{eq:fg_term} is computationally more expensive than evaluating the force $-\boldsymbol{e}_i(\mathcal{S})T^i$. 
Secondly, it requires the implementation of second-order derivatives contracted with first-order ones, which becomes non-trivial to implement, making the force-gradient approach impractical.

\section{Hessian-free force-gradient integrators}
One can overcome the aforementioned drawbacks of the force-gradient approach by approximating the force-gradient step \eqref{eq:fg-step}, as proposed in \cite{wisdom1996symplectic} for enhancing the Strang splitting \cite{strang1968construction}, and initially applied in the context of lattice QCD for a particular force-gradient integrator in \cite{yin2011LAT}. 
For a general force-gradient integrator, the approximation to the force-gradient step \eqref{eq:fg-step} reads \cite{schaefers2024hessian}
\begin{align}\label{eq:approximation_of_FGstep}
    \begin{split}
    \exp\!\left(b_j h \hat{\mathcal{B}} + c_j h^3 \hat{\mathcal{C}}\right) &=
    \exp\left(-b_j h \boldsymbol{e}_i(\mathcal{S}) \frac{\partial}{\partial p_i} + 2 c_j h^3 \boldsymbol{e}^j(\mathcal{S})  \boldsymbol{e}_j\boldsymbol{e}_i(\mathcal{S}) \frac{\partial}{\partial p_i} \right), \\
    &= \exp\left(-b_j h \left(\mathrm{Id} - \frac{2c_j h^2}{b_j} \boldsymbol{e}^j(\mathcal{S}) \boldsymbol{e}_j \right) \boldsymbol{e}_i(\mathcal{S}) \frac{\partial}{\partial p_i}\right), \\ 
    &= \exp\left(-b_j h \exp\left(- \frac{2c_j h^2}{b_j} F^j \boldsymbol{e}_j \right) \boldsymbol{e}_i(\mathcal{S}) \frac{\partial}{\partial p_i} \right) + \mathcal{O}(h^5),
\end{split}
\end{align}
where $F^j \boldsymbol{e}_j = \boldsymbol{e}^j(\mathcal{S})(U) \boldsymbol{e}_j$
 is regarded as a frozen vector field, i.e., $\boldsymbol{e}_j$ acting on $F^j$ is defined to be zero.
Since
$$\exp\left( -\tfrac{2c_j h^2}{b_j} F^j\boldsymbol{e}_j \right) \boldsymbol{e}_i(\mathcal{S})(U) = \boldsymbol{e}_i(\mathcal{S})\left( \exp\left(-\tfrac{2c_j h^2}{b_j} F^j T_j \right)U \right),$$
this approximated force-gradient step can be computed via the two-step procedure: 
\begin{enumerate}
    \item[1.] Compute a temporary link update via $U' = \exp\left( - \frac{2c_j h^2}{b_j} F^jT_j \right)U$;
    \item[2.] Compute a momentum update $P - b_j h \boldsymbol{e}_i(\mathcal{S})(U')T^i$.
\end{enumerate}
Consequently, the approximation replaces the force-gradient term with a second force evaluation, thereby reducing computational cost and rendering the framework accessible to existing software packages.
By replacing the force-gradient steps \eqref{eq:fg-step} with the approximation \eqref{eq:approximation_of_FGstep} that we denote by
\begin{equation}\label{eq:approximated_FG-step}
    \exp(b_j h \hat{\mathcal{D}}(h,b_j,c_j)),
\end{equation}
we obtain a new class of \emph{Hessian-free force-gradient integrators} \cite{schaefers2024hessian} that can be expressed as
\begin{equation}\label{eq:Hessian-free_FGI}
    \Phi_h =\mathrm{e}^{b_s h \hat{\mathcal{D}}(h,b_s,c_s)} \mathrm{e}^{a_s h \hat{\mathcal{A}}} \cdots \mathrm{e}^{b_1 h \hat{\mathcal{D}}(h,b_1,c_1)} \mathrm{e}^{a_1 h \hat{\mathcal{A}}}.
\end{equation}
In summary, Hessian-free force-gradient integrators utilize the force-gradient approach to enhance the computational efficiency without necessitating the Hessian $\boldsymbol{e}_j \boldsymbol{e}_i(\mathcal{S})$ of the action.

\subsection{Order conditions}
Similar to conventional splitting methods and force-gradient integrators, the order conditions are derived using the Baker--Campbell--Hausdorff (BCH) formula.
Due to the approximation of the force-gradient step, additional order conditions are obtained.
Since the approximation to the force-gradient step introduces an error of order $\mathcal{O}(h^5)$, the order conditions of Hessian-free and exact force-gradient integrators up to order $p=4$ are identical (for the order conditions, see~\cite{omelyan2003symplectic}).
In addition to the four order-6 conditions $\gamma_i = 0\; (i=1,2,3,4)$ stated in \cite{omelyan2003symplectic}, we obtain a fifth condition $\gamma_5 = 0$ stated in \cite{schaefers2024hessian}.
For order $p=8$, the ten conditions $\zeta_i = 0 \; (i=1,\ldots,10)$ of force-gradient integrators \cite{omelyan2003symplectic} are extended by three additional conditions $\zeta_i = 0 \; (i=11,12,13)$ \cite{schaefers2024hessian}.

\subsection{Geometric properties}
One disadvantage of the Hessian-free variants is their loss of symplecticity as the approximated force-gradient steps are no longer symplectic~\cite{hairer2009simplified,schaefers2024hessian}.
Since the approximated force-gradient steps \eqref{eq:approximated_FG-step} are shears, and a composition of shears is volume-preserving, Hessian-free force-gradient integrators remain volume-preserving. Consequently, they still satisfy the detailed balance condition, provided that the composition of the flows is self-adjoint.

Good energy conservation is crucial to ensure a high acceptance probability in the HMC algorithm. The preservation of a nearby shadow Hamiltonian \cite{kennedy2013shadow} guarantees good energy conservation. However, due to the absence of symplecticity, Hessian-free force-gradient integrators no longer preserve a nearby shadow Hamiltonian. Through a backward error analysis, it has been shown that a Hessian-free force-gradient integrator of order $p$ preserves the shadow Hamiltonian that is preserved by the underlying force-gradient integrator, along with a (in general linear) energy drift of size $\mathcal{O}(\tau h^{\max\{4,p\}})$ using the step size $h$ over a trajectory of length $\tau$ \cite{schaefers2024hessian}. 
This energy drift may pose a challenge for exponentially long-time simulations. In the HMC algorithm, however, $\tau$ is typically small, making the energy drift negligible and not observable in numerical tests.

\section{Derivation of efficient integrators}
Given a particular decomposition algorithm with a fixed number of stages, one can determine the maximum possible convergence order $p$.
For many variants, this results in a family of solutions where certain integrator coefficients remain as degrees of freedom.
A crucial aspect of deriving efficient numerical integration schemes involves appropriately selecting the degrees of freedom. 

\subsection{Minimum-error methods}
One potential approach to minimize the principal error term is to set all brackets equal to one, assuming that all error terms in the leading error term contribute equally. Subsequently, the norm of the leading error coefficients is minimized \cite{mclachlan1995,omelyan2003symplectic}. 
An approach that also addresses the computational cost is provided by the efficiency measure \cite{omelyan2003symplectic}
\begin{equation}
    \mathrm{Eff}^{(p)} = \frac{1}{(n_f + c \cdot n_g)^p \cdot \mathrm{Err}_{p+1}}.
\end{equation}
Here, $n_f$ and $n_g$ denote the number of force and force-gradient evaluations per time step, respectively, and $c$ a constant that mirrors the factor in computational cost of a force-gradient evaluation compared to a force evaluation. Moreover, $p$ denotes the convergence order of the method and $\mathrm{Err}_{p+1}$ some norm of the leading error terms. 
For the investigations of force-gradient integrators \cite{omelyan2003symplectic}, it was assumed that a force-gradient evaluation is twice as expensive as a force evaluation ($c=2$) and for $\mathrm{Err}_{p+1}$, the usual Euclidean norm of the leading error coefficients has been employed. 
In the Hessian-free framework, force-gradient evaluations are replaced by another force evaluation ($c=1$) and the norm has been extended by the additional error terms. For further details, see~\cite{schaefers2024hessian}.
The efficiency measure emerges as a valuable heuristic in identifying promising integrators within the family of decomposition algorithms.
However, in practical computations, it is observed that the integrator with the highest efficiency measure may not necessarily be the most efficient integrator.
This observation can be attributed to two primary reasons: Firstly, the norm of the leading error coefficients $\mathrm{Err}_{p+1}$ assumes that all error terms are equally significant. 
Secondly, decomposition algorithms are only conditionally stable. In the region of interest where \mbox{$0.01 \le \sigma^2(\Delta H) \le 1$}, the integrators frequently approach the boundary of their stability domain.
A highly accurate integrator may become impractical in practice if its stability domain is too small, especially for smaller fermion masses.
Therefore, an investigation of the numerical stability is paramount in evaluating the performance of the integrators.

\subsection{Linear stability analysis}
For splitting methods, a linear stability analysis is already available \cite{blanes2008stability} that considers the harmonic oscillator $\ddot{y} = -\omega^2 y$, $\omega > 0$, as a test problem. 
Decomposition algorithms typically will be unstable for $\lvert h \omega\rvert > z_*$, where the parameter $z_*$ denotes the stability threshold of the integrator.
Relying on the hypothesis for interacting field theories \cite{edwards1997instabilities,joo2000instability}, the high frequency modes of an asymptotically free field theory can be considered as a collection of weakly coupled oscillator modes. Consequently, the instability described in the harmonic oscillator will also be present for interacting field theories. Particularly, the onset of instability will be caused by the mode with highest frequency $\omega_{\max}$.
This motivates the extension of the linear stability analysis to (Hessian-free) force-gradient integrators. 
As the harmonic oscillator is a linear ODE system, force-gradient integrators and their Hessian-free variants are equivalent, i.e., their linear stability analysis coincides. 
By extending the linear stability analysis from \cite{blanes2008stability} to force-gradient integrators, one is able to compute the stability threshold $z_*$ for any force-gradient integrator so that the integrator is stable in the stability interval $(-z_*,z_*)$. 
To incorporate the computational cost of the integrators, it is worth considering the relative stability threshold $z_*/(n_f + c \cdot n_g)$ rather than the stability threshold itself.
Details on determining $z_*$ can be found in \cite{blanes2008stability} for splitting methods and are currently under preparation for (Hessian-free) force-gradient integrators \cite{preparation}.

\section{Selection of promising integrators}
\begin{figure}[tbh]
    \centering
    \includegraphics[scale=.925]{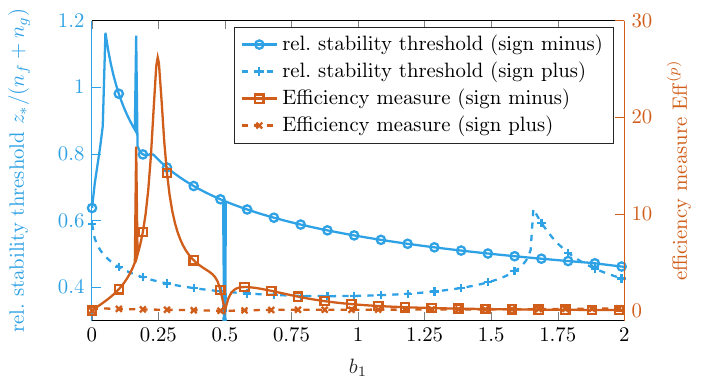}
    \caption{Analysis of the Hessian-free force-gradient integrator ABADABA in terms of the efficiency measure $\mathrm{Eff}^{(p)}$ and the relative stability threshold $z_*/(n_f + n_g)$.}
    \label{fig:ABADABA_analysis}
\end{figure}

In general, maximizing the efficiency measure and maximizing the stability threshold will result in different integrator coefficients. 
Initial observations emphasize that the efficiency measure $\mathrm{Eff}^{(p)}$ is way more sensitive to changes in the integrator coefficients than the stability threshold $z_*$.
As an example, Fig.~\ref{fig:ABADABA_analysis} shows results for the Hessian-free force-gradient integrator ABADABA
\begin{align*}
    \Phi_h &= \mathrm{e}^{a_1 h \hat{\mathcal{A}}} \mathrm{e}^{b_1 h \hat{\mathcal{B}}} \mathrm{e}^{(0.5 - a_1)h \hat{\mathcal{A}}} \mathrm{e}^{(1-2b_1)h \hat{\mathcal{D}}(h,1-2b_1,c_2)} \mathrm{e}^{(0.5 - a_1)h \hat{\mathcal{A}}} \mathrm{e}^{b_1 h \hat{\mathcal{B}}} \mathrm{e}^{a_1 h \hat{\mathcal{A}}}
\end{align*}
with integrator coefficients
\begin{align*}
    a_1 &= \frac{1}{2} \pm \frac{1}{\sqrt{24b_1}}, \qquad c_2 = \frac{1}{12} \left( 1 \pm \sqrt{6b_1}(1-b_1) \right), 
\end{align*}  
which is of convergence order $p=4$ with one degree of freedom $b_1 > 0$.
The results suggest that preference should be given to the negative sign in the formulas for $a_1$ and $c_2$.
When maximizing the efficiency measure, one still obtains $67\%$ of the maximum possible value for $z_*/(n_f+n_g)$. 
In contrast, maximizing $z_*/(n_f+n_g)$ yields only $3.9\%$ of the maximum possible value for $\mathrm{Eff}^{(p)}$. 
Instead, it is more promising to choose $b_1 = 1/6$ resulting in the integrator BADAB that demands one force evaluation less per time step, as can be seen by the peak in Fig.~\ref{fig:ABADABA_analysis}.
We observe similar results for other variants of decomposition algorithms. Hence we propose to keep maximizing the efficiency measure and then selecting those variants that have a reasonably large stability domain.
In Table \ref{tab:promising_integrators}, we summarize Hessian-free force-gradient integrators and non-gradient algorithms of order $p=4$ that are not dominated\footnote{An integrator is dominated, if another variant has a higher value for both the efficiency measure and the relative stability threshold.} by other variants.
\renewcommand{\arraystretch}{1}
\begin{table}[tbh]
    \centering
    \begin{tabular}{@{}l l l l l l l@{}}
        \toprule
        ID & $n_f$ & $n_g$ & $z_*$ & $z_*/(n_f + n_g)$ & $\mathrm{Err}_{p+1}$ & $\mathrm{Eff}^{(p)}$  \\\midrule
        BADAB & 2 & 1 & 3.4641 & 1.1547 & 0.000728 & 17.0 \\
        ABADABA & 3 & 1 & 3.1377 & 0.7844 & 0.0000149 & 26.2 \\
        BABABABABAB & 5 & 0 & 3.1421 & 0.6284 & 0.0000270 & 59.3 \\
        BADABADAB & 4 & 2 & 3.1457 & 0.5243 & 0.0000105 & 73.5 \\
        ABADABADABA & 5 & 2 & 3.1239 & 0.4463 & 0.00000445 & 93.6 \\\bottomrule
    \end{tabular}
    \caption{Self-adjoint Hessian-free force-gradient integrators and non-gradient schemes (with maximized efficiency measure) of order $p=4$ and with up to eleven stages that are not dominated by other variants. For the definitions and integrator coefficients of the respective schemes, see \cite{schaefers2024hessian}.}
    \label{tab:promising_integrators}
\end{table}

\section{Numerical Results}
\begin{figure}[H]
    \centering
    \includegraphics[scale=.875]{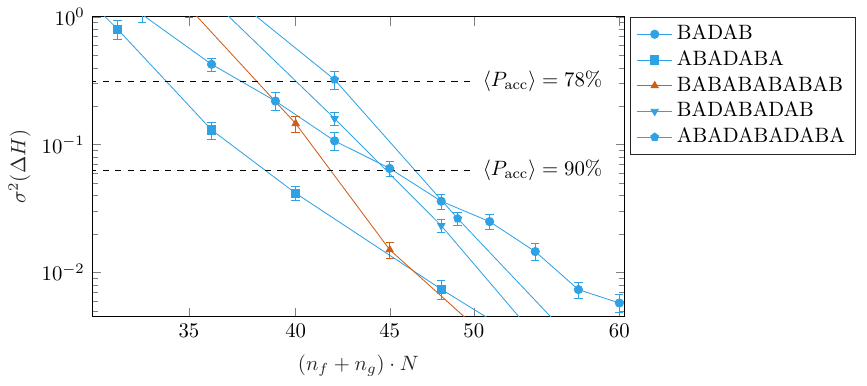}
    \caption{Variance of $\Delta H$ vs.\ number of force evaluations per trajectory $n_f \cdot N$ for all integrators from Tab.~\ref{tab:promising_integrators}. Here, the simulations have been performed with trajectory length $\tau = 2$.}
    \label{fig:Em1_results}
\end{figure}

\begin{figure}[tbh]
    \centering
    \includegraphics[scale=.875]{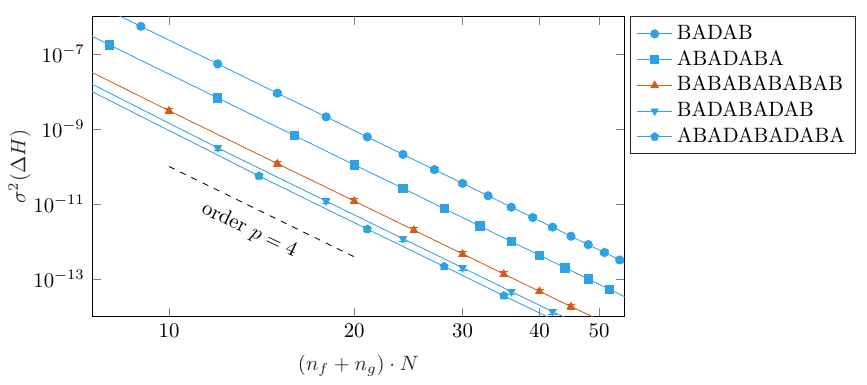} 
    \caption{Variance of $\Delta H$ vs.\ number of force evaluations per trajectory $n_f \cdot N$ for all integrators from Tab.~\ref{tab:promising_integrators}. Here, the simulations have been performed with trajectory length $\tau = 0.1$.}
    \label{fig:Em1_results_scaling}
\end{figure}

As an initial test of Hessian-free force-gradient integrators in lattice QCD simulations, we consider an ensemble used in \cite{knechtli2022} generated with two dynamical non-perturbatively $\mathcal{O}(a)$ improved Wilson quarks at a mass equal to half of the physical charm quark.
The fermion part is decomposed using even-odd reduction in combination with one Hasenbusch mass preconditioning term \cite{hasenbusch2001} with shift parameter $\mu$. For a more detailed discussion, we refer to \cite{frommer2014}.
On a lattice of size $48 \times 24^3$ with gauge coupling $\beta = 5.3$, hopping parameter $\kappa = 0.1327$, we performed numerical simulations using an extended version of \texttt{openQCD} v2.4\footnote{\href{https://github.com/KevinSchaefers/openQCD_force-gradient}{https://github.com/KevinSchaefers/openQCD\_force-gradient}, based on \href{https://luscher.web.cern.ch/luscher/openQCD/}{openQCD v2.4}.} 
by putting all forces on a single time scale of integration. Particularly, it turns out that at these lattice parameters, the use of nested integration techniques \cite{urbach2006} employing a smaller step size to the gauge part does not result in significant improvements in the acceptance probability.
Starting from a thermalized configuration, we computed 100 trajectories of length $\tau=2$ for varying step sizes $h=\tau/N$ ($N \in \mathbb{N}$) for all integrators from Tab.~\ref{tab:promising_integrators}. The results are depicted in Fig.~\ref{fig:Em1_results}, demonstrating that the integrator ABADABA allows for the most efficient computational process, despite its lower value for the efficiency measure. This highlights the importance of the relative stability threshold already for larger fermion masses.\newline
In a second simulation with smaller trajectory length $\tau = 0.1$, resulting in smaller time steps $h=\tau/N$, the integrators are no longer affected by numerical instabilities, as depicted in Fig.~\ref{fig:Em1_results_scaling}. In this "scaling phase", the performance of the integrators perfectly matches the efficiency measure.\newline
For the best-performing Hessian-free force-gradient integrator ABADABA and the best-performing non-gradient scheme BABABABABAB, we tuned the step size to achieve an acceptance rate of $P_{\mathrm{acc}} \geq 90\%$, resulting in $h=0.2$ for both integrators, and computed 2000 trajectories of length $\tau = 2$. The results for both setups are summarized in Tab.~\ref{tab:tuning}, including the cost measure $\mathtt{cost} = (n_f+n_g) \cdot N/(P_{\mathrm{acc}} \cdot \tau)$. 
On the one hand, we obtain similar estimates of the integrated autocorrelation times of the topological charge $Q_0$ measured at Wilson flow time $t_0$ \cite{Luscher:2010iy}, and the Wilson flow reference scale $t_0/a^2$, despite the lower acceptance rate of ABADABA. 
On the other hand, the cost measure indicates that the Hessian-free force-gradient integrator only demands approximately $84\%$ of the computational cost compared to the best-performing non-gradient scheme.

\renewcommand{\arraystretch}{1}
\begin{table}[tbh]
    \centering
    \begin{tabular}{@{}l l l l l l@{}}
         \toprule
         ID & $(n_f + n_g) \cdot N$ & $P_{\mathrm{acc}}$ & $\tau_{\mathrm{int}}(t_0)$ [MDU] & $\tau_{\mathrm{int}}(Q_0)$ [MDU] & $\mathtt{cost}$ \\\midrule
         BABABABABAB & 50 & 97.5\% & 37.40(12.66) & 22.91(6.52) & 25.64 \\
         ABADABA & 40 & 92.3\% & 28.15(9.24) & 22.05(6.31) & 21.67 \\\bottomrule 
    \end{tabular}
    \caption{Comparison of two tuned setups based on 2000 trajectories of length $\tau = 2$. Both setups use $10$ time steps per trajectory ($h=0.2$). The results contain the number of force evaluations per trajectory $(n_f +n_g) \cdot N$, the acceptance rate $P_{\mathrm{acc}}$, estimates for the integrated autocorrelation times $\tau_{\mathrm{int}}(t_0)$ and $\tau_{\mathrm{int}}(Q_0)$ in molecular dynamics units (MDUs), and an evaluation of the cost measure.}
    \label{tab:tuning}
\end{table}
\vspace*{-3ex}

\section{Conclusion}
Hessian-free force-gradient integrators constitute a promising choice as integrators for performing lattice QCD simulations. 
They utilize the force-gradient approach to enhance the computational efficiency without necessitating the Hessian of the action that is part of the force-gradient term. 
As volume-preserving and time-reversible integrators, these integrators satisfy the detailed balance condition. 
The approximation of the force-gradient step results in additional order conditions and the loss of symplecticity. 
Numerical results demonstrate that neither the additional error terms nor the lack of symplecticity have a significant impact on the energy conservation of the integrator. As a conclusion, the Hessian-free framework provides more efficient integrators and can be easily integrated into existing software packages.

\acknowledgments
This work is supported by the German Research Foundation (DFG) research unit FOR5269 "Future methods for studying confined gluons in QCD". The simulations were carried out on the PLEIADES cluster at the University of Wuppertal, which was supported by the DFG and the Federal Ministry of Education and Research (BMBF), Germany.
J.F.~acknowledges financial support by the Next Generation Triggers project (\href{https://nextgentriggers.web.cern.ch}{https://nextgentriggers.web.cern.ch}).

\bibliographystyle{JHEP}
\bibliography{mybib}

\end{document}